\documentstyle[12pt]{article}
\input epsf

\begin{document}

\begin{titlepage}
\begin{center}
\null\vskip-1truecm
\vskip2.2truecm {\bf SPIN PUZZLE IN
NUCLEON\footnote{\normalsize Internal Report, ICTP, MIRAMARE, Trieste,
Italy : IC/94/261, September 1994}}
\vskip1.2truecm
R. Ramachandran \\
Institute of Mathematical Sciences, Madras, India \\
\end{center}
\vskip1.5truecm

\centerline{ABSTRACT}
\bigskip

The object of this brief review is to reconcile different points of view on
how the spin of proton is made up from its constituents. On the basis of
naive quark model with flavour symmetry such as isospin or SU(3) one
finds a static description. On the contrary the local SU(3) colour
symmetry gives a dynamical view. Both these views are contrasted and
the role of U(1) axial anomaly and the ambiguity for the measurable spin
content is discussed.

 \vfill
\end{titlepage}

\section*{Introduction}

Matter in the universe is mostly made up of nucleons (and as many
electrons as there are protons). It is important that the nucleons are spin
1/2 particles and the concommitant Fermi statistics is responsible for
the fact that many hard objects can be made of these constituents. When it
became clear that the nucleons are not elementary and they are
presumably made up of other spin 1/2 species known as quarks, it is useful
to understand how the nucleon spin is distributed in its constituents.

There are two different SU(3) groups associated with the strong nuclear
interactions. The flavour SU(3) group is the improvement on the older
isospin symmetry indicated by the near equality of masses of proton and
neutron. Adding strangeness as an additional flavour, in the early sixties,
the underlying symmetry governing the properties of nucleons was
established to be SU(3). Soon it became clear that the neutron and proton
are two members of an octet of baryons (other members of the family
being an isosinglet $\Lambda_0$, isotriplet $\Sigma^+,\Sigma^0$,
$\Sigma^-$ both carrying strangeness -1 and an isodoublet $\Xi^{0,-}$ with
-2 units of strangeness) and the triplet of pions are accompanied by the
doublets of $K(K^+,K^0)$ and $\bar K(\bar K^0,K^-)$ mesons and an
isosinglet $\eta$ meson, thus making up an octet of spin zero odd parity
(pseudo scalar) mesons. That neither the set of baryons nor mesons have
same mass for all its member states is an indication of the approximate
nature of the symmetry. This symmetry has been useful in labelling the
states and describes, along with the energy, momentum and angular
momentum the kinematical state. The other SU(3), believed to be exact,
governs the dynamics of the strong interactions. The strong interaction
dynamics is similar to the more familiar quantum electrodynamics (QED)
described by the local interactions that has gauge invariance and the
related conservation of electric charge as the underlying principle. The
minimal electromagnetic coupling is obtained by the simple substitution
of the energy-momentum operator $p_\mu(={\hbar\over
i}\partial_\mu)$ by the canonical momentum
$p_\mu-eA_\mu(={\hbar\over i}\partial_\mu-eA_\mu)$ in the Lagrangian.

The local symmetry of gauge invariance (the transformation that lets the
wavefunction of a charged particle $\psi(x)\to
e^{i\alpha(x)Q_{em}}\psi(x)$;
$Q_{em}$ is the electromagnetic charge operator) is ensured by endowing
the gauge field $A_\mu(x)$ representing the photon, the property that it
transforms as $A_\mu(x)\to A_\mu(x)-\partial_\mu\alpha(x)$. The gauge
transformations $\psi(x)\to e^{i\alpha(x)Q}\psi(x)$ and $A_\mu(x)\to
A_\mu(x)-\partial_\mu\alpha(x)$ leaves the Lagrangean invariant. The
symmetry group parametrized by $\alpha$ is $U(1)$, elements of unitary
matrices of unit dimension -- just the phases of a unimodular complex
number -- an abelian group of transformations and is related is the
conservation of the electromagnetic charge.
The strongly interacting matter carries analogous, but non-abelian, charge
characterised by the unitary unimodular group SU(3), the generators of the
group being exp(i$\alpha_aQ_a)$; $a=1,2,\dots 8$. The symmetry
transformation will be an operator $U(x)=\exp[i\alpha_a(x)Q_a]$;
$\psi(x)\to U(x)\psi(x)$. The local symmetry is ensured by endowing now
the matrix valued gauge field $A_\mu(x)(\equiv A^a_\mu(x)Q^a)$ the
property that it transforms as $A_\mu(x)\to UA_\mu
U^{-1}-U\partial_\mu U^{-1}$. Like QED, the quantum chromodynamics
(QCD) is a renormalizable field theory that lets us obtain finite
regularized matrix elements for all physically relevant amplitudes.

It is not surprising that kinematic and dynamic SU(3) groups are
both relevant in the description of nucleons (and other baryons and
mesons) as composite states of quarks, which possess ``charges'' that are
consequences of both symmetry groups. The spin content of the nucleon, I
will argue, arises out of differing roles of the kinematic SU(3), that
specifies the flavour quantum numbers of hadronic matter and the
dynamic SU(3), related to the ``colour'' of the strongly interacting matter.

Nucleon and other baryons are bound states of three quarks; however
unlike the atomic states, there is no concept of ``ionisation'' or ``
dissociation'' of quarks, since the states carrying colour quantum
numbers are not found as proper asymptotic states. This is referred to as
the property of confinement of colour-- a property of the QCD, the precise
mechanism for which is yet to be clarified. While it is impossible to pull
quarks out of the nucleon, at very short distances probed by high
momentum-transfer photons, the quarks appear to be ``weakly'' interacting,
thereby justifying the use of perturbative quantum chromodynamics for
the description of deep inelastic scattering of leptons by nucleons. We
will contrast the spin structure of the nucleon revealed in such a study
with what one may expect on the basis of the wavefunction of a static
nucleon in terms of its flavour and spin content.

\section*{Flavour SU(3) and Static Spin Structure of Nucleons}

To begin with, it has been popular in the sixties to view nucleons as a three
body bound state [1], while two {\it up} quarks and one {\it down} quark
 make up a proton, two $d$ and one $u$ will constitute a neutron. It was
possible to accomodate an octet of baryons and a decimet of $J^P=3/2^+$
excited states in a 56 dimensional representation of SU(6) symmetry
group\footnote{\normalsize The up, down and strange quarks with $\pm
1/2$ spin states each makes up the six dimensional fundamental
representation of SU(6) symmetry.}.
The three-quark symmetric state spans a 56 dimensional representation,
made up of $J=1/2$ doublets of baryon octet and $J=3/2$ spin quartet of
excited baryon decimet. The latter is made up of the prominent $p$-wave
meson-baryon resonances:
$\Delta(1232$ MeV, $I=3/2$, $S=0$), $\Sigma^*(1395$ MeV, $I=1$,
$S=-1$) and $\Xi^*$ (1530 MeV, $I={1\over 2}$, $S=-2$) and the stable
(with respect to strong interactions) state $\Omega^-(1672$ MeV,
$I =0$, $S=-3$). The proton wave function (with spin
$s_z=1/2)$ is given in terms of the quarks through [2]: (for neutron wave
function interchange
$u$ and $d$ quarks)
\begin{eqnarray}
\psi_p & = & \{{1\over\sqrt{18}}\{2\vert u^+_1 d^-_2u^+_3\rangle
+2\vert u^+_1u^+_2d^-_3\rangle +2\vert d^-_1 u^+_2
u^+_3\rangle\nonumber\\
&& - \vert u^+_1u^-_2 d^+_3\rangle -\vert u^+_1 d^+_2 u^-_3\rangle
-\vert u^-_1d^+_2u^+_3\rangle\nonumber\\
&& - \vert d^+_1u^-_2u^+_3\rangle -\vert d^+_1 u^+_2u^-_3\rangle -\vert
u^-_1u^+_2 d^+_3\rangle\}
\end{eqnarray}
where the letters $u$ and $d$ refers to up and down quarks $\pm$ to spin
$s_z=\pm1/2$ and the subscripts 1, 2 and 3 to label the position
(space-time $\chi^\mu_i$; $i= 1,2,3$) of the quarks. In such a quark model
all relative orbital angular momenta vanish in the ground state and thus
the proton spin is carried symmetrically in terms of the constituent quark
spin. Further the static magnetic dipole moments of the states are given
by the operator
\begin{equation}
\vec\mu_B =\mathop{\Sigma}\limits_q \mu_q\vec\sigma_q
\end{equation}
We may read off the expectation value for $\mu_p$ and $\mu_n$:
$$\mu_p ={4\over 3}\mu_u-{1\over 3}\mu_d\ .\eqno(3a)$$
$$\mu_n = {4\over 3}\mu_d-{1\over 3}\mu_u\eqno(3b)$$

Since the magnetic dipole moment and angular momentum are related
through
$$ \vec\mu =g\mu_0\vec j;\quad \mu_0={e\hbar\over 2mc}\ {\rm and}\
g=-2\ {\rm for\ Dirac\ fermion}\ .$$
We expect for each quark $$\mu_q=Q_q\left({e\hbar\over
2m_qc}\right)\eqno(4)$$
where $Q_q$ is the fractional unit of charge carried by the quark $q$ and
$m_q$ is the ``constituent'' quark mass.

Thus
$\mu_n={2\over 3}\left({m_p\over m_u}\right) \mu_N,\quad
\mu_d=-{1\over 3}\left({m_p\over m_d}\right)\mu_N$ and $\mu_s
=-{1\over 3}\left({m_p\over m_s}\right)\mu_N$, where $\mu_N$ is the
Nucleon Bohr magneton. Isospin symmetry will lead to $m_u=m_d=\bar m$
so that
$$\mu_p={m_p\over m} \mu_N =2.79 \mu_N\Rightarrow \bar m=336\ {\rm
MeV}\eqno(5)$$
With this, we should get
$$\mu_n = -{2\over 3}{m_p\over \bar m}\mu_N=-1.86 \mu_N\eqno(6)$$
as against the experimental value of $\mu_n=-1.91 \mu_N$.
Indeed the magnetic moments of all baryonic states ($p, n,
\Lambda,\Sigma,\Xi$ as well as $\Delta,\Sigma^*,\Xi^*$ and $\Omega^-$)
are given in terms of just two parameters $(\bar m=m_u=m_d$ and $m_s$);
say $\mu_p$ and $\mu_\Lambda$. Experimental numbers match the static
quark model prediction to a large extent.
$$
\begin{tabular}{lll}
Magnetic moment & Quark model values & Experimental values
$(in\ \mu_N)$\\
\\
$\mu_p$ & ${4\over 3}\mu_u-{1\over 3}\mu_d\ ({\rm input})$ & 2.793\\
$\mu_n$ & ${4\over 3}\mu_d-{1\over 3}\mu_u=-1.86\mu_N$ & $-1.913$\\
$\mu_\Lambda$ & $\mu_s ({\rm input})$ & $-0.613\pm 0.004$\\
$\mu_{\Sigma^+}$ & ${4\over 3}\mu_u -{1\over 3} \mu_s=2.69 \mu_N$ &
$2.458\pm0.010$\\
$\mu_{\Sigma^0-\Lambda}$ & ${1\over \sqrt 3}
(\mu_u-\mu_d)=1.65\mu_N$ & $1.61 \pm 0.08$\\
$\mu_{\Sigma^-}$ &${4\over 3}\mu_d-{1\over 3}\mu_s =-1.04 \mu_N$
&$-1.160\pm 0.025$\\
$\mu_\equiv$&${4\over 3}\mu_s-{1\over
3}\mu_u=-1.44\mu_N$&$-1.250\pm0.014$\\
$\mu\equiv^-$&${4\over 3}\mu_s-{1\over
3}\mu_d=-0.51\mu_N$&$-0.679\pm 0.031$\\
$\mu_{\Omega^-}$&$3\mu_s=-1.84\mu_N$&$-1.94\pm0.22$
\end{tabular}
$$

\centerline{\bf Table 1. Magnetic Moments of Baryons}
\bigskip

 We may conclude that the
naive quark model describes adequately the static spin structure of the
nucleon.

\section*{Colour SU(3) and dynamical spin content}

The spin content of the nucleon is probed dynamically by the longitudinal
polarization asymmetries in the deep inelastic scattering by leptons off
the nucleon targets [3]. The matrix elements of both electromagnetic and
weak currents provide valuable information as the structure of the
hadrons. Typically for $\ell(k)+p(p)\to \ell'(k')+X$, we have, in the leading
order, the contribution
for the inclusive cross section is given by the product of the leptonic
tensor $\ell_{\mu\nu}(=k^m_\mu k'_\nu+k'_\mu k_\nu-g_{\mu\nu}(k\cdot
k')+\in_{\mu\nu\alpha\rho}m s^\alpha k^\rho)$ and the hadronic tensor
$$W^{\mu\nu}={1\over 2\pi}\int d^4 x\exp(iq\cdot x)\langle p, s\vert
[j^\mu(x),j^\nu(0)]\vert p,s\rangle$$
and the $W^\pm,Z$ or $\gamma$ propagator.

While the symmetric part of $W_{\mu\nu}$ measures the unpolarized
hadronic structure, the spin content is coded in the antisymmetric part.
The Lorentz covariance of the hadronic tensor together with the
constraint arising from the conservation of electromagnetic current will
imply for $e+N\to e+X$, two structure functions $F_1 \ \mbox{and} \ F_2$
for the symmetric part (and
these are analogous to the electric and magnetic form factors of the
elastic scattering) and two more $g_1 \ \mbox{and} \ g_2$ for the antisymmetric
part:
\setcounter{equation}{6}
\begin{eqnarray}
{M\over 2\pi}  W^{\mu\nu} & = &
(g^{\mu\nu}-q^\mu q^\nu/q^2)F_1(q^2,\nu\equiv p\cdot q/2M)\nonumber\\
&& +{1\over M\nu} \left(p^\mu -{(p\cdot q)q^\mu\over
q^2}\right)\left(p^\nu-{(p\cdot q)q^\nu\over q^2}\right)
F_2(q^2,\nu)\nonumber\\
&& -{1\over \nu} \in^{\mu\nu\alpha\beta}q_\alpha s_\beta
g_1(q^2,\nu)-{1\over\nu}\in^{\mu\nu\alpha\beta}q_\alpha(s_\beta-{q\cdot
s_p)\over p\cdot q}) g_2(\nu,q^2)
\end{eqnarray}
where $q^2(=-Q^2)$ and $\nu$, the energy transfer in the lab frame are
related to the two independent kinematic variables (energy transfer and
the scattering angle of the lepton) that can be measured in an all inclusive
deep inelastic scattering process. It is instructive to   use instead, the
variables $Q^2(\equiv -q^2=4 k_0k'_0\sin^2\theta/2)$ and $x(\equiv
Q^2/2p\cdot q =Q^2/2M\nu)$  and find that the structure function have only
a very weak (logarithmic) dependence on $Q^2$ and is mostly expressible in
terms
of the scaling variable $x$. In the quark-parton model, one views the
process of deep inelastic scattering as an incoherent sum of the
scattering of the lepton by the quark that carries a momentum fraction $X$
of the parent nucleon and the scattering event with the variable $X$ is
correlated with the quark carrying corresponding fractional momentum
$X= x$. Thus the measured structure functions are directly translated as
the probabilities of quark distributions in a hadron. What is more, the
slow logarithmic $Q^2$ dependence -- the so called scaling violation --
can be computed in the perturbative quantum chromodynamics and turns
out to be the important evidence for QCD as the experimentally
verifiable strong interaction theory
[4].

It is easy to derive in the leading order,
$$2x F_1(x)=F_2(x)=x[\mathop{\Sigma}\limits_i
e^2_i[(q^+_i(x,Q^2)+q^-_i(x,Q^2))
+(q^+_{\bar i}(x,Q^2)+q^-_{\bar i}(x,Q^2))]
\eqno(8)$$
$$\equiv \Sigma_i xe^2_i[q_i(x,Q^2)+q_{\bar i}(x,Q^2)]$$
where $q_i(x,Q^2) = q^+_i + q^-_i$ is the net probability of finding a quark of
flavour $i$,
carrying a  momentum of fraction $x$ of the parent proton. The summation
will run over all quark and antiquark flavours. The superscript $\pm$
refers to the helicity of the quark inside, say, a positive helicity proton.
Perturbative QCD gives the $Q^2$ evolution of the various moments of the
quark densities and the experimental verification of the expected scaling
violation (i.e. $\ell n Q^2$ dependence of the $q^n_i(Q^2)$ ($= \int^1_0 dx
x^{n-1}q(x,Q^2)))$ is hailed as the triumph of QCD. The first part of Eq.(8),
relating $F_1(x)$ and $F_2(x)$ is a consequence of the spin 1/2 nature of
quarks.

The spin structure is similarly revealed in the functions $g_1(x,Q^2)$
and $g_2(x,Q^2)$. For the longitudinal asymmetries $g_2(x,Q^2)$ does not
contribute -- since $s^\sigma$ is parallel to $p^\sigma$. In the leading
order,
\setcounter{equation}{8}
\begin{eqnarray}
g_1(x,Q^2) & = & {1\over
2}\Sigma_ie_i^2[(q^+_{\bar
i}(x,Q^2)-q^-_ i(x,Q^2))+q^+_{\bar i}(x,Q^2)-q^-_{\bar
i}(x,Q^2))]\nonumber\\ &
\equiv & {1\over 2} \Sigma_i e_i^2[\Delta q_i(x,Q^2)+\Delta
q_{\bar i}(x,Q^2)]
\end{eqnarray}
The $Q^2$ dependence is governed by the Altarelli-Parisi equation [5],
which has a very transparent interpretation as a process of finding within
a parton, (quark of flavour $i$ or gluon $G$) another parton (quark of
flavour $j$ or gluon $G$) and is expressed as a convolution of a splitting
function $p(z)$, $z$ being the momentum fraction. The splitting functions
$p(z)$ are given in terms of the basic vertices of QCD and the resultant
integro-differential equation can be solved to give ordinary algebraic
differential equations for the moments of quark and gluon densities. For
non-singlet flavour (all valence quarks) quark densities, we get
$(q^{valence}_i \equiv q_i-q_{\bar i})$
$$ {d\over dt} \Delta q^{valence,n}_i (t) ={\alpha_s(t)\over 2\pi} \tilde
A^n_{qq} \Delta q^{valence,n}_i(t)\eqno(10)$$
where $t=\ell n Q^2/Q^2_0$; $Q^2_0$ being a reference scale,
$\alpha_s(t)$ $(=g^2_s/4\pi)$ is the running coupling parameter in QCD
(analogue of $\alpha_{em}$) with $Q^2$ dependence given by
$$\alpha_s(t)=\alpha_s(0)/(1+b\alpha_s(0)t);\quad b = (33-2f)/12\pi\
.\eqno(11)$$
for the colour group SU(3) with $f$ flavours. $\tilde A^n_{qq}$ are the
anomalous dimensions of the relevant operator and can be read off the QCD
splitting functions. $\alpha_s(t)$ decreases as $t$ increases and signifies
the asymptotic freedom of quark gluon coupling at short distances,
permitting therefore the  perturbative QCD analysis. We find:
$$\Delta
q^{valence,n}_i(t) =[\alpha_s(0)/\alpha_s(t)]^{\tilde A^n_{qq}/2\pi b}\Delta
q^{valence}_i(0)\eqno(12)$$
For $\Delta q^{valence,1}_i$, which measures the spin structure in terms
of the difference of number of quarks with $+$ and $-$ helicities, the
solution  is particularly simple, since $\tilde A^1_{qq}=0$ and so is
$t$-independent. If we identify these valence quark densities with the
SU(6) wave function (Eq. 1 and 3a,b), we have [8]
$$\Delta q^{valence,1}_u=4/3,\quad \Delta q^{valence,1}_d=-1/3\
.\eqno(13)$$
together with the unwanted consequence of SU(6) symmetry, viz.
$$ G_A/G_V\bigg\vert_{p\to n} =\Delta q^{v,1}_u-\Delta
q^{v,1}_d={5\over 3}\ .\eqno(14)$$
As an alternate to the pure SU(6) description it is also possible to use
the experimental value of
$$G_A/G_V \bigg\vert_{p\to n}=1.25 =\Delta
q^{v,1}_u-\Delta q^{v,1}_d$$
 and a similar relation for
$$G_A/G_V\bigg\vert_{\Xi^-\to\Xi^0}=-0.25=\Delta q^{v,1}_d$$
 to give [7]
$$\Delta q^{v,1}_u=1\ {\rm and}\ \Delta q^{v,1}_d=-0.25\eqno(15)$$
Yet another option is to use the model of Carlitz and Kaur, who propose
that the valence quark hypothesis and most of the momentum and helicity
is carried by the ``leading quark'' in the hadron [8]. Accordingly
$$\Delta q^v_u(x) = \cos 2\Theta[q^v_u(x)-{2\over 3}q^v_d(x)]$$
$$\Delta q^v_d(x)=-{1\over 3}\cos^2\Theta (x) q^v_d(x)\eqno(16)$$
with $\cos 2\Theta (x) = [1+0.052(1-x^2)/\sqrt
x]^{-1}$.

Since $\cos 2\Theta (x)$, called spin-dilution factor, approaches unity as
$x\to 1$ and since $q^v_u(x)$ dominates over $q^v_d(x)$ at large $x$,
the valence $u$-quarks in protons have their spins aligned fully along
proton spin in this kinematical region. The first moment -- the net number
of valence spin -- is remarkably close to the values in Eq. (15), since the
integral of Eq.(16) yields:
$$\Delta q^{v,1}_u = 1.01\ {\rm and}\ \Delta
q^{v,1}_d=-0.25\eqno(17)$$

\section*{Flavour Singlet Spin Puzzle}

The nucleon has, in addition to the valence quarks, $q\bar q$ pairs referred
to as the sea and gluons\footnote{\normalsize We ignore the possibility that
it may also have a very small component of heavy quarks charm, bottom or
top.}. It was generally believed that both the sea and gluons are unpolarized
and the spin is carried entirely by the valence quarks. It is remarkable that
this is far from the truth and there is a strong tendency for the quarks and
gluons to be polarized. This was conjectured by me through an analysis of
the $Q^2$ evolution in the early eighties [9] and subsequent experimental
measurement [10] of EMC (European Muon Collaboration) gave values for
longitudinal spin asymmetry that implied that the bulk of the spin is
resident elsewhere. Closely related to this feature is the notion that QCD
has U(1) axial anomaly and as a consequence what is being perceived as
the quark spin (in the flavour singlet part) contains a strong component of
gluon spin, which will normally be ignored as being higher order in
$\alpha_s(Q^2)$. Denote $G^+(x,Q^2)-G^-(x,Q^2)=\Delta G(x,Q^2)$, where
$\pm$ refer to gluon helicities. The U(1) axial anomaly causes the term
$\alpha_s(Q^2)\Delta G^1(Q^2)$ to be $Q^2$ independent [11] and it gets
added to the flavour singlet quark spin $\Sigma_i\Delta q^1_i(Q^2)$. Thus
the flavour singlet spin content of the hadron is measured to be
$\Sigma_i[\Delta q^1_i(Q^2)-{\alpha_s(Q^2)\over 2\pi}\Delta G^1(Q^2)]$.

The QCD evolution equation for the singlet sector is given by [12]
$${d\over dt}\left(\begin{array}{c}
\Sigma_i\Delta q^n_i(t)\\
\Delta G^n(t)\end{array}\right) =(\alpha_s(t)/2\pi)
\left(
\begin{array}{cc}
\tilde A^n_{qq} & 2f\tilde A^n_{qq}\\
\tilde A^n_{Gq}&\tilde A^n_{GG}
\end{array}\right)
\left(
\begin{array}{c}
\Sigma_i\Delta q^n_i\\
\Delta G^n
\end{array}\right)\eqno(18)$$
$\tilde A^n$ are known constants related to the anomalous dimensions of
relevant operators, obtainable from the basic QCD couplings. For $n=1$,
they are rather special with $\tilde A^1_{qq}=0=\tilde A^1_{qG}$, $\tilde
A^1_{Gq}=3 c_F$ and $\tilde A^1_{GG}$ $(={11c_A-4T\over 6})=2\pi b$
where $c_F$ (Casimir invariant for the fundamental representation),
$c_A$ (Casimir invariant for the adjoint representations) have  values, for
4/3 and 3, respectively when the gauge group is $SU(3)$ \ and $T=f/2$.
$${d\over dt}\mathop{\Sigma}\limits_i\Delta q^1_i(t)=0\eqno(19a)$$
and
$${d\over dt}\Delta G^1(t) = (\alpha_s(t)/2\pi)(2\Sigma_i \Delta
q^1_i+2\pi b \Delta G^1)\eqno(19b)$$
To this, we must add the helicity sum rule
$${1\over 2} \Sigma_i \Delta q^1_i+\Delta G(t)+L_z(t)={1\over 2}\
.\eqno(20)$$
Since the renormalized gauge coupling constant satisfies ${d\over
dt}\alpha_s(t)=-b\alpha^2_s(t)$, it is easily observed that
$${d^2\over dt^2}\Delta G^1=0\ {\rm upto}\ {\cal O}(\alpha^2_s)$$
Thus, the right hand side of Eq. (19b) gives
$${\alpha_s(t)\over\pi}(\Sigma_i\Delta q^1_i+\pi b\Delta
G^1)=C\eqno(21)$$
where $C$ is a constant ($t$ independent). Indeed the helicity sum rule
(Eq.20) calls for
$$\langle L_z(t)\rangle =\langle L_z(0)\rangle -Ct\eqno(22)$$
with the same constant $C$. All the above equations are consistent with
the following values for the first moments:
$$\Sigma\Delta q^1 =\left({33-2f\over 9-2t}\right)
(1-2(L_z(0)\rangle)\eqno(23a)$$
and
$$\Delta G^1=\left({-12\over 9-2f}\right)
(1-2L_z(0)\rangle)+Ct\eqno(23b)$$
Equation (23a,b) indicate non-trivial magnitude of the proton spin to be
found in the favour singlet part. It may also be observed, as per the
helicity sum rule, while $\Sigma \Delta q^1_i$ is conserved quantity
$\Delta G^1$ and $L_z$ are not. This is easily understood as a general
property that the fermion helicity is conserved in the QCD process, leaving
the emitted gluon to have either helicity and the total angular momentum
conservation will so adjust to have the net orbital angular momentum
$L_z$ and $\Delta G^1$ together remain conserved. These features are
reflected in Eqs. (22) and (23a,b).

The most significant feature is that $\alpha_s(Q^2)\Delta G^1(Q^2)$ is the
conserved quantity and in an experiment that involves measuring the
singlet spin, this quantity is not distinguished from the net quark spin.
Most straightforward way to see this is to look at the axial vector U(1)
current and observe that it is not conserved due to the presence of the
anomaly.
$$j^5_\mu =\Sigma^f_{i=1} \bar q_i\gamma_\mu \gamma_5q_i\eqno(24)$$
In the massless (for quarks) limit, the axial anomaly gives
$$\partial^\mu j^5_\mu = f{\alpha_s\over 2\pi} Tr(F_{\mu\nu}\tilde
F^{\mu\nu})\eqno(25)$$
where $\tilde F^{\mu\nu}={1\over
2}\in^{\mu\nu\rho\sigma}F_{\rho\sigma}$,
$F_{\mu\nu}=\Sigma^8_{a=1}F^a_{\mu\nu}Q^a$; $Q^a$: generators of
$SU(3)_e$ and $trQ^aQ^b={1\over 2}\delta^{ab}$. The right hand side is
expressible as a four-divergence of a current $k_\mu$:
$$k_\mu=\left({\alpha_s\over 2\pi}\right)
\in_{\mu\nu\lambda\sigma}Tr[A^\nu(F^{\lambda\sigma}-{2\over
3}A^\lambda A^\sigma)]$$
so that now the Eq. 25 is reexpressed as
$$\partial^\mu(j^5_\mu-fk_\mu) = 0\ .\eqno(26)$$
The conserved quantity $(\Sigma \Delta q^1_i - {f\alpha_s\over
2\pi}\Delta G^1)$ may be regarded as a consequence of the above
conservation law. Strictly speaking $k_\mu$, being gauge dependent,
should not occur as as a physically measurable quantity. However it turns
out that the diagonal matrix elements of this operator is gauge
independent and in a ``parton model'' like analysis one deals only with
diagonal matrix elements.

Altarelli and Lampe [14] observe that it will be appropriate to develop the
$t$-evolution equation for the two combination of densities.
Denoting $\Delta \Gamma ={\alpha_s\over 2\pi}\Delta G^1$ and $\Delta
\Sigma=\Sigma_i\Delta q^1_i$, upto two loop level, we have
$${d\over dt}\Delta\Gamma =2\left({\alpha_s\over 2\pi}\right)^2 (\Delta
\Gamma)\eqno(27)$$
and $${d\over dt}\Delta\Sigma = 0\ .\eqno(28)$$
It is seen that the quantity $\Delta\Sigma'\equiv (\Delta\Sigma -f\Delta
\Gamma)$ satisfies the two loop evolution equation:
$${d\over dt}\Delta \Sigma' =-2f\left({\alpha_s\over 2\pi}\right)^2
\Delta\Sigma'\eqno(29)$$
and can be solved to give
$$\Delta \Sigma' (t) =\Delta\Sigma' (\mu^2)\exp-2f\int
\left({\alpha_s\over 2\pi}\right)^2 dt\eqno(30)$$
$$\cong \Delta\Sigma'(\mu^2)\exp({f\alpha_s(t)\over \pi b})$$
if we neglect higher order terms in $\alpha_s(t)$\footnote{\normalsize We
should have used
$\alpha_s(t) \sim {1\over bt} +{b_1\over 2b^2t^2}$ with
$b_1={306-38f\over 48\pi^2}$].}.
 It is seen that $Q^2$ dependence is rather
weak and may be neglected to begin with in any phenomenological analysis.

\section*{Phenomenological Consequences}

The EMC measurement at CERN of the longitudinal asymmetry in the deep
inelastic muon-proton scattering yielded directly $g^p_1(x,Q^2)$. At
$Q^2=10.7 GeV^2$, it was found to give the first moment, with suitable
extrapolation, yielding [10]
$$\int^1_0g^p_1(x,Q^2)=0.126\pm 0.010+0.015\eqno(31)$$
Since this is the measure of ${1\over 2}\Sigma e^2_i\Delta q^1_i(Q^2)$, it
is seen to be the combination ${1\over 12}(\Delta q^1_u -\Delta
q^1_d)+{1\over 36}(\Delta q^1_u+\Delta q^1_d -2\Delta q^1_s)+{1\over
8}(\Delta q^1_u+\Delta q^1_d+\Delta q^1_s)$. Using the information on
$g_A/g_V(=\Delta q^1_u-\Delta q^1_d)$ and the $D/F$ radio for octet
baryon matrix element, extract the singlet part to give
$$\Sigma\Delta q^1_i=0.00\pm 0.24\eqno(32)$$
This result is indeed  the puzzle. This gave rise to the notion that
the flavour singlet spin structure is vanishingly small and
needed an explanation.
 The analysis made by Altarelli and Stirling [11] makes use of
the EMC result, $G_A/G_V\bigg\vert_{p\to n}$ and the $D/F$ ratios
as determined by the weak interaction of the octet of baryon
$\beta$-decays to give $(\Delta q^1_u-\Delta q^1_d=G_A/ G_V=1.25$;
with $D=0.79$ and $F=0.46$
$\Delta q^1_u+\Delta q^1_d-2\Delta q^1_s=3F-D=0.59)$,
$$\Delta q^1_u-\Delta \Gamma =0.74\pm 0.08$$
$$\Delta q^1_d-\Delta\Gamma =-0.51\pm 0.08$$
$$\Delta q^1_s-\Delta \Gamma =-0.23\pm 0.08\eqno(33)$$
and rewrite the Eq.32  as
$$\Sigma_i \Delta q^1_i-3\Delta\Gamma-0.00\pm 0.24\eqno(32a)$$
$\Delta\Gamma=\alpha_s/2\pi \Delta G^1$ is the anomaly contribution to
quark spin in proton. If we take $\Delta q^1_s=0$, then
$\Delta\Gamma=0.23$ and $\Sigma\Delta q^1_i=3\Delta\Gamma=0.69$,
which implies 30\% of the proton spin is due to the gluon. The fraction of
spin carried by quarks decreases rapidly for negative $\Delta q^1_s$. For
$\Delta q^1_s =-0.10, \ \mbox{we will have} \ \Delta \Gamma =0.13\Rightarrow
\Sigma\Delta
q^1_i=0.39$, which then means 60\% of the proton spin is carried by the
gluon component. These numbers are indicative of a not too dramatic
depletion of quark spin content in proton.

Subsequent experiments have improved the quality of the data and the spin
Muon Collaboration (SMC) now reports data that are much less dramatic.
The results reported at the Glasgow conference [15] suggests
$$\Sigma\Delta q^1_i\cong 0.31\pm 0.07$$
and the violation of Ellis-Jaffe sum rule, interpreted as the strangeness
content of proton spin gives
$$\Delta q^1_s\cong -0.10\pm 0.04$$

Notwithstanding the absence of dramatically puzzling experimental
result, the spin puzzle cannot be regarded as a solved problem. What we
have seen in this note is the inherent ambiguity in what is theoretically
analyzed and what is amenable to measurement. The theoretical analysis
makes it clear that what is measured as the singlet component of the spin
and for that matter any flavoured quark spin is the combination consisting
of the naive parton model quark  spin and $-{\alpha_s\over 2\pi}\Delta G$
for each flavour. Since this combination occurs in the (only) conserved
current of the theory, it is quite clearly  what is measured. This is the
reason, we believe, following Altarelli and Lampe, that what is
perceived as quark spin in experiments is the combination $(\Delta
q^1_i-\alpha_s(Q^2)\Delta G^1(Q^2)/2\pi)$, which show $Q^2$ evolution
with terms of order ${\cal O}(\alpha^2_s)$ and higher.

Gluon polarization can be probed by studying large $p_T$ $pp$ scattering
process in which a prompt photon is observed [16]. The dominant hard
scattering process, that gives rise to prompt photon is the compton
analogue; a valence quark of one proton scatters off a polarized gluon
$g+q\to\gamma+q$.
To probe the proton spin content in gluon, we may either use single
polarized proton to observe circular polarization of  the prompt photons
$(p+p^\uparrow\to\gamma^{\uparrow\downarrow}X)$ or study the
longitudinal polarization asymmetry in the scattering of polarized proton
on a polarized target
$(p^\uparrow+p^{\uparrow\downarrow}\to \gamma+X$). It is found that in
both these cases, the dominant contribution comes from the``compton''
subprocess and hence can be used a clear probe for the gluonic content of
the proton spin.

Before we conclude it may be worthwhile reminding us that the structure
functions are intrinsically non-perturbative inputs in QCD. The anal;ysis,
we have presented largely banks on the one hand on the perturbative QCD
and the effect it has on the {\it evolution} of the structure functions and
on the other hand as the consistency conditions that arise from the underlying
symmetries and conservation principles. It will be desirable to see what
one can say about the spin content from an ab-initio non-perturbative
model, say through a lattice computation. In view of the reported success
of the technology of lattice gauge computation in being able to get low
energy parameters, it should be possible to arrive at reliable results on
the spin structure as well.

\section*{Acknowledgments}

This review is based on the author's study with many associates
R.P. Bajpai, M. Rafat, M. Noman and P. Mathews over many years among
others.
The author would like to thank Professor Abdus Salam, the International
Atomic Energy Agency and UNESCO for hospitality at the International
Centre for Theoretical Physics, Trieste, and the congenial environment in
which this was written up.

\newpage

\section*{References}
\begin{enumerate}
\item
See for example M. Gell-Mann and Y. Ne'emen, {\it The Eightfold Way},
Benjamin, New York (1964);\\
F.E. Close, {\it An Introduction to Quarks and Partons}, Academic Press,
New York (1979).

\item
Fayyazuddin and Riazuddin, {\it A Modern Introduction to Particle
Physics}, World Scientific, Singapore (1992), Ch. 6.

\item
See for example, R.P. Feynman, {\it Photon-Hadron Interactions},
Benjamin, New York (1972); P. Roy, {\it Theory of Lepton-Hadron
Processes at High Energies}, Oxford University Press, Oxford (1975).

\item
G. Altarelli, Phys. Rep. {\bf 81C} (1982), 1.

\item
G. Altarelli and G. Parisi, Nucl. Phys. {\bf B126} (1977), 298.
\item
J. Kuti and V.F. Weisskopf, Phys. Rev. {\bf D4} (1971) 3418.

\item
L.M. Sehgal, Phys. Rev. {\bf D10} (1974) 1663.

\item
R. Carlitz and J. Kaur, Phys. Rev. Letters {\bf 88} (1977) 673, 1102;\\
J. Kaur, Nucl. Phys. {\bf B128} (1977) 219.

\item
R.P. Bajpai and R. Ramachandran, Phys. Lett. {\bf B97} (1980) 125;\\
R.P. Bajpai, M. Noman and R. Ramachandran, Phys. Rev. {\bf D24} (1981)
1832.

\item
J. Ashmon et al. (EMC Collaboration) Phys. Lett. {\bf B206} (1988) 364 and
J. Ashmon et al. Nucl. Phys. {\bf B328} (1989) 1.

\item
M. Rafat and R. Ramachandran, Pramana {\bf 23} (1984) 675.;\\
G. Altarelli and W.J. Stirling {\it Particle World} {\bf 1} (1989) 40.

\item
R. Ramachandran, Pramana {\bf 21} (1983) 11.

\item G. Altarelli and G. Ross, Phys. Lett. {\bf B212} (1988) 391;\\
A.V. Efremov and O.V. Teryaev, Dubra preprint E2-88-287 (1988).

\item
G. Altarelli and B. Lampe, Zeit. f. Physik {\bf C47} (1990) 315.

\item
J. Feltesse, Rapporteur's talk at ICHEP-94 at Glasgow (to be published).

\item
P. Mathews and R. Ramachandran, Zeit. f. Physik. {\bf  C53} (1992) 305.

\item D. Indumathi, M.V.N. Murthy and S. Gupta, Zeit. f. Physik {\bf C47}
(1990)
227.
\end{enumerate}
\newpage

\begin{center}

\section*{Figure Captions}

{\bf Figure 1} - Deep  inelastic scattering to probe the dynamical spin
content of nucleon.

\vskip 1.5in

\hspace*{\fill}
\hspace*{\fill}

\end{center}
\end{document}